\documentclass{amia}
\usepackage{graphicx}
\usepackage[labelfont=bf]{caption}
\usepackage[superscript,nomove]{cite}
\usepackage{color}
\usepackage{subcaption}
\usepackage{amsfonts, textcomp}
\usepackage{xcolor}
\usepackage{graphicx}
\usepackage{url}
\usepackage[symbol]{footmisc}

\begin{document}

\title{Brain Atlas Guided Attention U-Net for White Matter Hyperintensity Segmentation}

\author{Zicong Zhang, MS$^{1}$, Kimerly Powell, PhD$^{2, 3}$,
        Changchang Yin, MS$^{1}$, Shilei Cao, MS$^{4}$, Dani Gonzalez$^{5}$,
         Yousef Hannawi, MD$^{6,*}$,  Ping Zhang, PhD, FAMIA$^{1, 2, *}$}
\institutes{
    $^1$Computer Science and Engineering, The Ohio State University, Columbus, Ohio, USA\\
    $^2$Biomedical Informatics, The Ohio State University, Columbus, Ohio, USA\\
    $^3$Department of Radiology, The Ohio State University, Columbus, Ohio, USA\\
    $^4$Tencent Jarvis Lab,  Tencent, Shenzhen, China\\
    $^5$Biomedical Engineering, The Ohio State University, Columbus, Ohio, USA\\
    $^6$Department of Neurology, The Ohio State University, Columbus, Ohio, USA\\
    $^*$Corresponding authors: yousef.hannawi@osumc.edu; zhang.10631@osu.edu
}

\maketitle

\noindent{\bf Abstract}

\textit{White Matter Hyperintensities (WMH) are the most common manifestation of  cerebral small vessel disease (cSVD) on the brain MRI. Accurate WMH segmentation algorithms are important to determine cSVD burden and its clinical consequences.
Most of existing WMH segmentation algorithms require both fluid attenuated inversion recovery (FLAIR) images and T1-weighted images as inputs. However, T1-weighted images are typically not part of standard clinical scans which are acquired for patients with acute stroke. In this paper, we propose a novel brain atlas guided attention U-Net (BAGAU-Net) that leverages only FLAIR images with a spatially-registered white matter (WM) brain atlas to yield competitive WMH segmentation performance. Specifically, we designed a dual-path segmentation model with two novel connecting mechanisms, namely multi-input attention module (MAM) and attention fusion module (AFM) to fuse the information from two paths for accurate results. Experiments on two publicly available datasets show the effectiveness of the proposed BAGAU-Net. With only FLAIR images and WM brain atlas, BAGAU- Net outperforms the state-of-the-art method with T1-weighted images, paving the way for effective development of WMH segmentation. Availability: \url{https://github.com/Ericzhang1/BAGAU-Net}}

\section*{Introduction}
Cerebral Small Vessel Disease (cSVD) is a major public health burden leading to vascular cognitive impairment, intracerebral hemorrhage, and acute ischemic stroke \cite{cognitive5}. Histologically, the white matter in patients with cSVD exhibits areas of demyelination and pallor that are seen as white matter hyperintensities (WMH) on the brain magnetic resonance imaging (MRI) \cite{icbm1,elastix2}. A higher volume of WMH is thought to represent a higher burden of cSVD. Indeed, WMH volume load has been suggested as a potential biomarker for cSVD burden. Clinical studies have shown that a larger WMH volume are associated with cognitive impairment, worse stroke functional outcome, and worse response to acute stroke therapy \cite{cognitive,cognitive2,cognitive3,cognitive4}. 
Hence, practical methods for accurate segmentation of WMH to determine its volume are currently needed. Manual tracing WMH in MRI brain images is currently the accepted method for segmenting WMH on brain MRI images. However, it is time-consuming and may be associated inter-observer variability which limits its applicability in daily clinical practice and real time decision making in patients with acute stroke.

Deep neural networks have shown robust performance in WMH segmentation tasks compared to manual segmentation \cite{wmh-skip,sysu_media,stack-nets,location}. Most of these methods make use of T1-weighted and FLAIR imaging sequences acquired for each patient without considering their
clinical usage. In general, the T1-weighted image provides the spatial location information of white matter (WM) that helps guide the WMH segmentation process \cite{t1}. Additionally, research MRI scans are often acquired using standardized protocols in a homogenous group of patients that limit their applicability to routine clinical MRI scans. However, in the real-world clinical setting, rapid MRI acquisition in the setting of acute ischemic stroke without T1 sequences have become the normal protocol to allow for critical time sensitive clinical decision making in this setting \cite{t1_2,t1_3}.  
Hence, there is a critical need to develop a WMH segmentation method based on FLAIR sequences only that can be employed in the treatment of acute stroke patients.
Brain atlases have been developed in the past to assist in imaging segmentation by improving preprocessing and image registration and they have been utilized in WMH segmentation as well \cite{add1, add2}. However, these approaches heavily rely on the presence of T1 sequences for adequate registration and segmentation \cite{add3, add4}.
Recently, atlas-based segmentation was implemented in deep learning, which is either (1) employed in a neural network to learn the correspondences between target image and atlas images \cite{panet} or (2) used to provide guidance as prior knowledge during model training \cite{deepatlas}. Motivated by the latter case, we explored the use of publicly available WM brain atlas to guide our WMH segmentation in the absence of corresponding T1-imaging sequence.

In this paper, we propose a novel brain atlas guided attention U-Net (BAGAU-Net) for WMH segmentation to address the primary limitations mentioned above. BAGAU-Net consists of two segmentation paths that take the FLAIR image and the spatially registered WM brain atlas separately to provide accurate segmentation results. The two segmentation paths are combined using two novel attention-based connecting mechanisms.
Our contribution in this work is four fold: (1) We propose to use only FLAIR and publicly available WM brain atlas to achieve robust performance on WMH segmentation compared to using T1-weighted image; (2) we propose an end-to-end dual path model called brain atlas guided attention U-Net (BAGAU-Net) that leverage an additional path, namely atlas encoding path, to effectively capture the prior knowledge from WM brain atlas to help improve segmentation performance; (3) we introduce multi-input attention module (MAM) and attention fusion module (AFM) to combine the information from two paths to further improve the segmentation performance; (4) we evaluate our model on two publicly available datasets, the 2017 MICCAI WHM segmentation
challenge\footnote[1]{\url{https://wmh.isi.uu.nl}} and the Aging Brain: Vasculature, 
Ischemia, and Behavior Study (ABVIB) dataset\footnote[2]{\url{https://ida.loni.usc.edu/}}. The results show that BAGAU-Net has out-performed previously proposed state-out-the-art method on both datasets.

\section*{Method}

As aforementioned, T1-weighed images are not acquired frequently for WMH segmentation due to time constraints in most the clinical settings. In contrast, FLAIR images are expected in the treatment of acute stroke
patients, but does not possess detailed information as T1-weighted image. To relieve such a dilemma, we propose to explore extra prior knowledge to improve the performance with only FLAIR images to match the performance with T1-weighted images. Inspired by the classical concepts of atlas-based segmentation, we propose to exploit prior knowledge hidden in a WM brain to guide the segmentation of the FLAIR images. To this end, we introduce the BAGAU-Net architecture for WMH segmentation. We first describe the generation process of standard WM brain atlas. Then we go into details about the dual-path architecture, including (1) the segmentation path, (2) the atlas encoding path, (3) the multi-input attention module (MAM), and (4) the attention fusion module (AFM).


\begin{figure*}[!htbp]
\centering
\subcaptionbox{}{\includegraphics[width=0.24\textwidth, height=100pt]{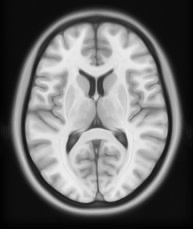}}%
\hfill 
\subcaptionbox{}{\includegraphics[width=0.24\textwidth, height=100pt]{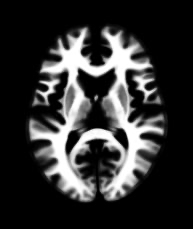}}%
\hfill
\subcaptionbox{}{\includegraphics[width=0.24\textwidth]{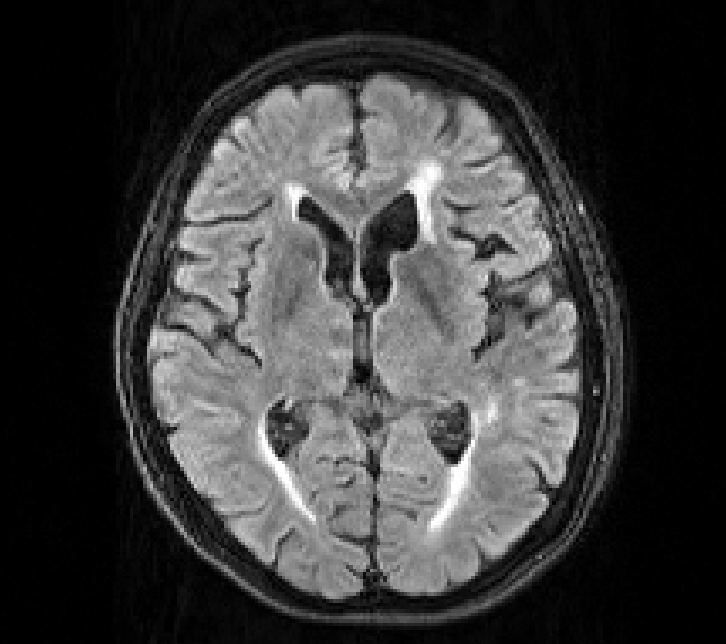}}%
\hfill 
\subcaptionbox{}{\includegraphics[width=0.24\textwidth]{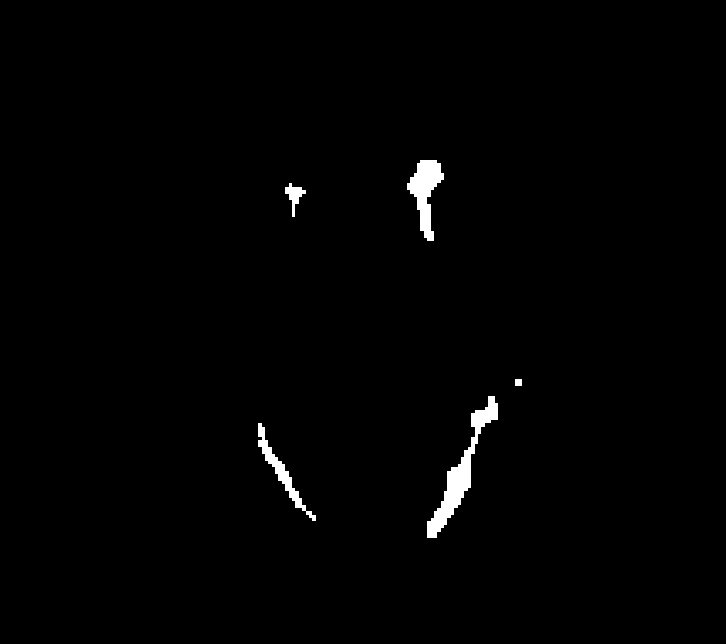}}%
\caption{Examples of MR images, from left to the right are: (a) T1-weighted ICBM152 atlas, (b) corresponding WM probability atlas, (c) FLAIR image, (d) manual WMH segmentation}
\label{teaser}
\end{figure*}

\subsection*{Atlas Generation}

Brain atlases are often used to identify neuroanatomical structures of the brain. We used 
the ICBM152 (2009c Nonlinear Symmetric, $1\times1\times1$ in 
mm)\footnote[3]{\url{http://www.bic.mni.mcgill.ca/ServicesAtlases/ICBM152NLin2009}} and the
corresponding WM brain atlas \cite{icbm1,icbm2} for our model. {\itshape Elastix} \cite{elastix1,elastix2} was used to spatially register the T1-weighted ICBM152 atlas to each of our target images with the following parameters: A multi-resolution pyramid with three levels using advanced mattes mutual information, standard gradient descent optimizer, maximum number of iterations 2,000, and a third order B-spline interpolator. The corresponding ICBM152 WM brain atlas was then transformed to match the target image. An example of spatially registered ICBM152 T1-weighted and WM brain atlas, target FLAIR and manually segmented WMH are shown in Fig. \ref{teaser}.

\subsection*{Model Architecture}

\begin{figure}[!htbp]
\includegraphics[width=1.0\textwidth]{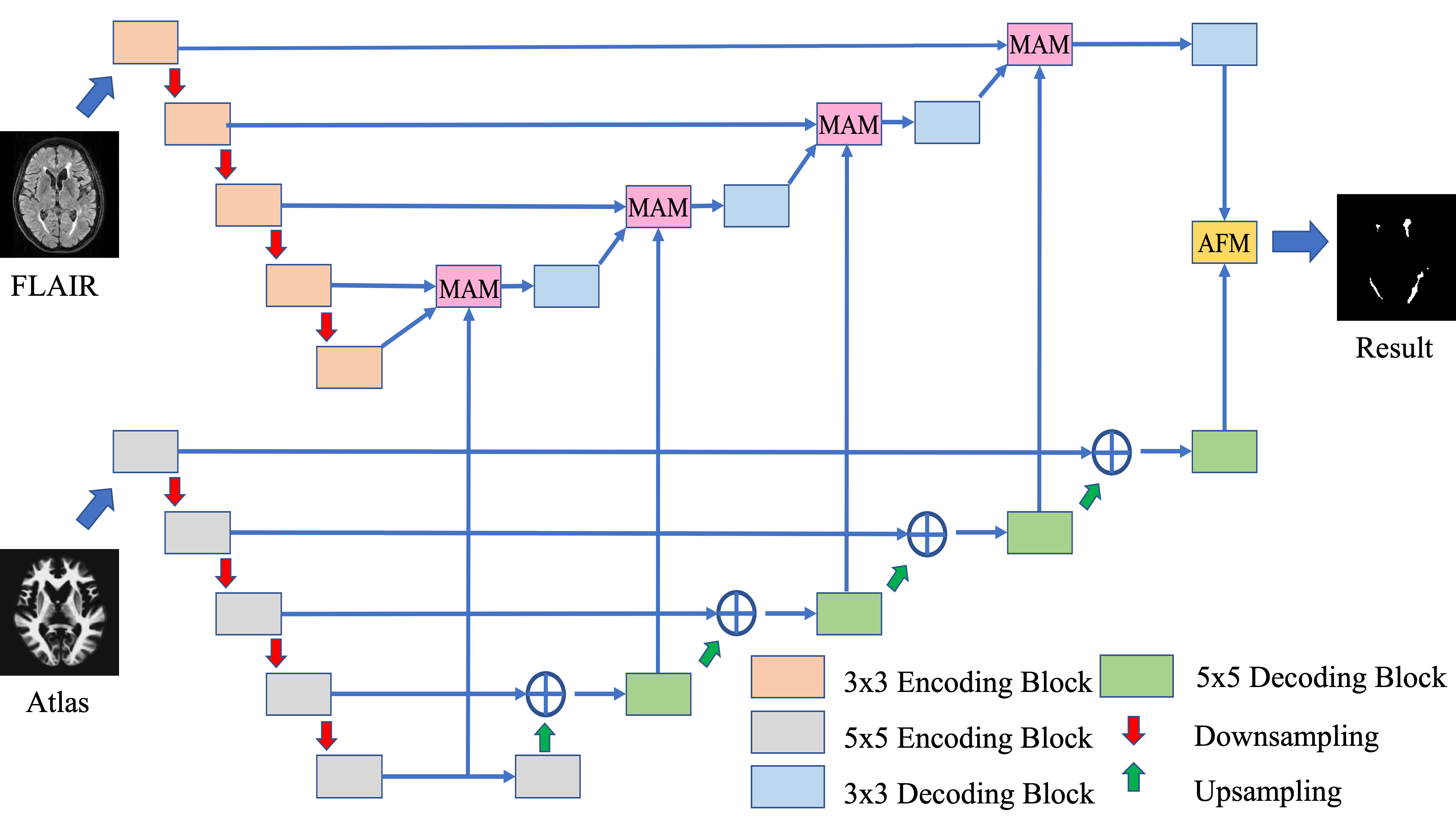}
\caption{Overview of the BAGAU-Net for WMH segmentation. The model consists of 3x3, 5x5 encoding and decoding block, multi-input attention module (MAM), attention fusion module (AFM), upsampling, and pooling operation.} \label{model}
\end{figure}

Convolution neural networks have shown robust performance in many segmentation tasks. As inspired by Li \textit{et al.} \cite{sysu_media}, we proposed brain atlas guided attention U-Net (BAGAU-Net) that consists of two separate encoding-decoding paths. As shown in Fig~\ref{model}, the upper path is a U-Net like architecture designed to extract semantic information from the image itself. The lower path is the atlas encoding path where the spatially registered atlas image is input to help guide the decoding process in the segmentation path. Moreover, we designed a multi-input attention module (MAM) and attention fusion module (AFM) to effectively combine the information between the two paths during the decoding process of segmentation path based on the attention gate (AG) as introduced by Oktay \textit{et al.} \cite{att-unet}. The overview of the proposed architecture can be viewed in Fig. \ref{model}.

\subsubsection*{Segmentation Path}

The segmentation path is designed to extract semantic features from the MRI scans. We adopt a U-Net like architecture that consists of four consecutive encoding blocks with feature channel equaling to $64, 96, 128, 256, 512$ followed by four decoding blocks connected by skip connections. Each encoding block consists of two consecutive 3$\times$3 convolutions followed by batch normalization and a rectified linear unit (ReLU). The max-pooling operation is applied for the down-sampling process to extract high-level features. Each decoding block are constructed in a similar fashion as the encoding architecture, except that an up-sampling operation is applied at the end of the block.

\subsubsection*{Atlas Encoding Path}

The atlas encoding path is used to encode spatial information from atlas images and serves as supplementary features during the decoding process of the segmentation path. We have adopted a similar architecture as our segmentation path with two modifications: (1) each 3$\times$3 convolution in substitute with 5$\times$5 convolution as inspired by Peng \textit{et al.} \cite{kernel} to capture multi-scale features. (2) we use addition instead of concatenation as inspired by Long \textit{et al.} \cite{FCN} to combine lower and higher level features.

\begin{figure*}[!t]
  \centering
  \begin{subfigure}[b]{0.48\textwidth}
    \includegraphics[width=\textwidth]{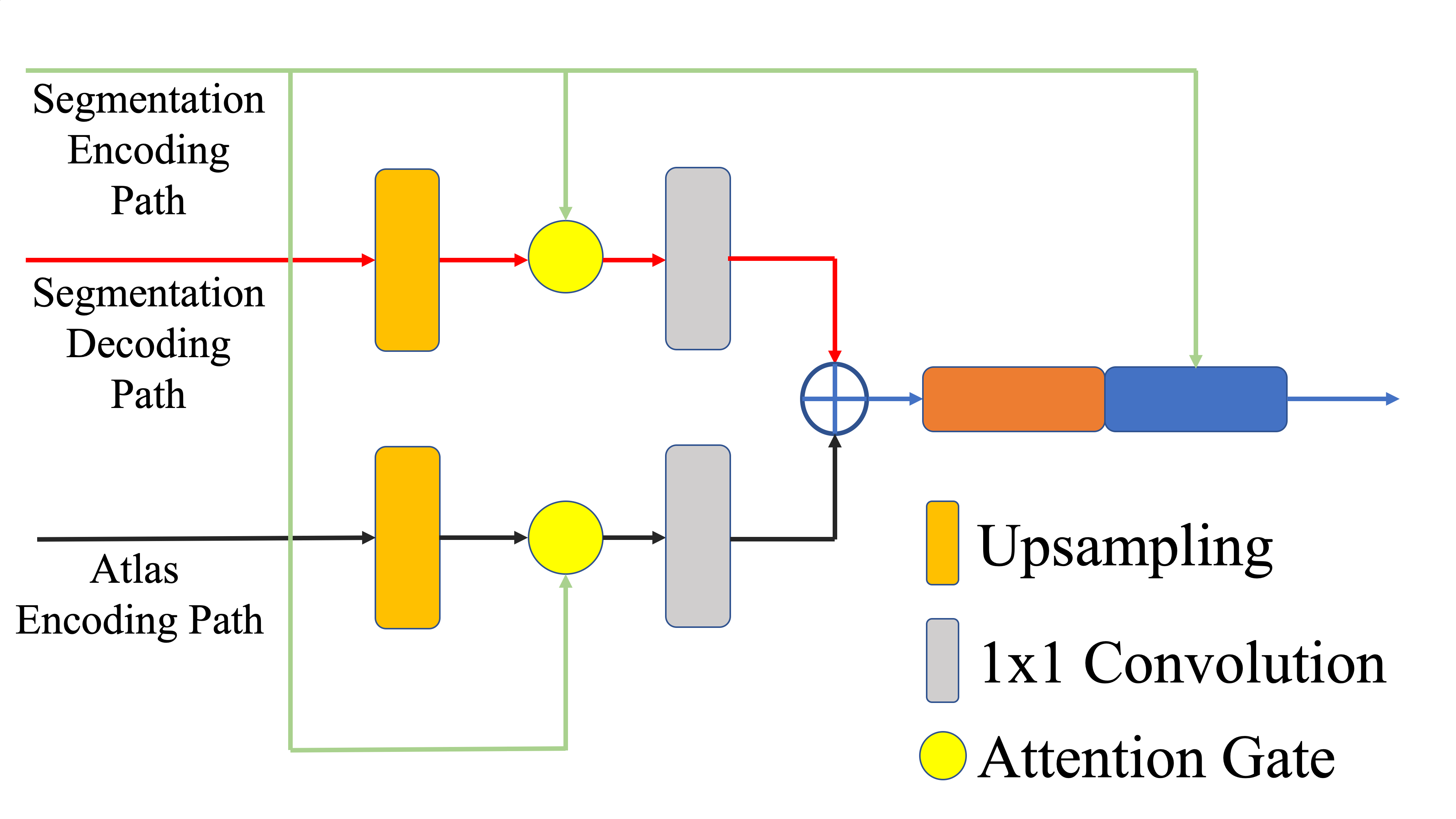}
    \caption{The structure of Multi-input Attention Module (MAM)}
    \label{MAM}
  \end{subfigure}
  \hfill
  \begin{subfigure}[b]{0.48\textwidth}
    \includegraphics[width=\textwidth]{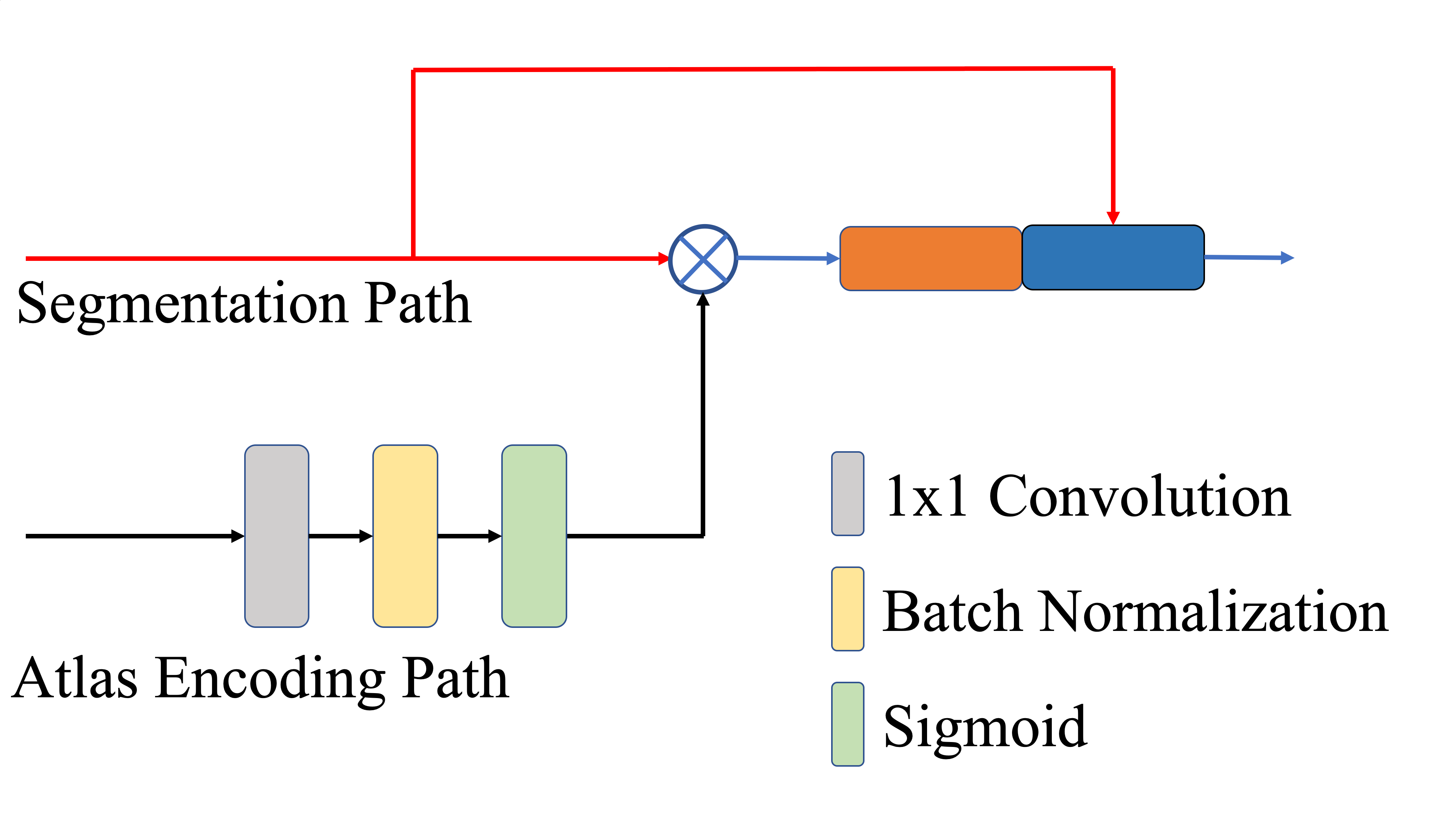}
    \caption{The structure of attention fusion module (AFM)}
    \label{AFM}
  \end{subfigure}
  \caption{The structure of components in BAGAU-Net}
  \label{components}
\end{figure*}

\subsubsection*{Multi-input Attention Module}

We implemented an attention mechanism that takes multi-input to compute target-aware features. Attention Gate, as proposed by Oktay \textit{et al.} \cite{att-unet}, is an efficient way of extracting salient features and contextual information, which allows the model to learn to focus on a subset of target features.
$g_i \in\mathbb{R}^{Fg}$ a gating vector and $x_i \in\mathbb{R}^{Fx}$ be the input to the attention gate. The attention gate computes the attention signal $\alpha_{i}\in[0, 1]$ using addition attention, which is as:

$$
   q_{att} =  W^{att}(\sigma(W_{x}^{T}x_{i} + W_{g}^{T}g_{i} + b_{g})) + b_{att},
$$
$$
   \alpha_{i} = \sigma(q_{att}(x_{i}, g_{i}; \theta_{att})).
$$

As shown in Fig. \ref{components} (a), MAM takes input from both segmentation path and atlas encoding path at each level and computes the combined feature based on the re-weighted summation of results computed by the attention gate.

\subsubsection*{Attention Fusion Module}

Given the output from the last layer of both the segmentation path and atlas encoding path, we implemented a channel-attention mechanism to refine the feature of the last layer. The output from the atlas encoding path is used to compute the attention mask, where, as shown in Fig. \ref{components} (b), the result is then fused with the output from the segmentation path using element-wise multiplication. A final convolution layer is applied to produce the segmentation result based the concatenation of the fused features.

\section*{Experiment}

\subsubsection*{Datasets}

The 2017 MICCAI WMH segmentation challenge dataset is publicly available. The dataset consists of 60 training subjects collection from 3 different types of scanners. A more detailed description of the challenge dataset can be found in Table. \ref{challenge_dataset}. For each subject, a FLAIR, T1, and the ground truth (manual segmentation of WMH) are provided. All images were bias-corrected using SPM12. We performed image registration through {\itshape Elastix}. The ABVIB dataset is a publicly available dataset originally established to study aging brains and cognitive consequences secondary to cardiovascular risk factors. 30 FLAIR subject images were selected and bias-corrected without associated 

\begin{table}[!htbp]
\centering
\caption{Overview of the 2017 MICCAI WMH segmentation challenge dataset}
\label{challenge_dataset}

\begin{tabular}{lcc}
\hline
Institute & Scanner & Size \\
\hline
UMC Utrecht & 3 T Philips Achieva & 20 \\
NUHS Singapore & 3 T Siemens TrioTim & 20 \\
VU Amsterdam & 3 T GE Signa HDxt & 20 \\
\hline
\end{tabular}
\end{table}

T1 sequences. Manual segmentation of WMH for ground truth was performed by a trained student and independently verified by two investigators including a board certified neurointensivist and neurologist.
 
\subsubsection*{Model Training and Implementation Details}
We split both WMH challenge and ABVIB dataset into training set, validation set and testing set with a ratio of 8:1:1. Instead of using DSC loss for segmentation, we adopt the Tversky loss metric \cite{tversky_loss} for all model training to provide more balanced weighting between false positives (FPs) and false negatives (FNs). Let P and G be the predict and true label correspondingly. The Tversky loss can expressed as:

$$
    T(P, G; \alpha) = \frac{|PG|}{|PG| + \alpha|P \setminus G| + (1 - \alpha)|G \setminus P|},
$$

where $\alpha$ in the above equation is a hyper-parameter that controls the trade off between FP and FN. We set the value of $\alpha$ to 0.7 based on the results of extensive search by Salehi \textit{et al.} \cite{tversky_loss}. We use Adam optimizer with learning rate of 0.0002 and batch size of 32 for all models; the number of epochs is set to 200. To demonstrate the effectiveness of the atlas encoding path, we compared our proposed model to the winning team's solution \cite{sysu_media} as well as the same architecture but simply adding the atlas knowledge as a third channel input.
We implemented our model using Pytorch platform \cite{pytorch} and trained the model on single Nvidia Volta V100 GPU with 16GB memory. We adopt gradient accumulation when dealing with out of memory problems. We employed data augmentation to both datasets, including mirroring, rotation, shearing and scaling. Gaussian normalization is further applied in addition to bias-correction to reduce variation across images. A threshold of $0.5$ was applied to transform the output of the model into a binary segmentation map.

\subsubsection*{Model Evaluation}
The results of the BAGAU-Net segmentation were evaluated both by visual comparison of the automated segmentation and the manually segmentations and by using quantitative metrics. Four quantitative metrics were used for the quantitative assessment: Dice coefficient (DSC), average volume difference (AVD), Recall, and F1 score were calculated for each segmentation. Dice coefficient (DSC) is a measure of the amount of overlap between two segmented structures. The average volume difference measures the total volume difference between ground truth and the predicted segmentations. Recall and F1 were calculated and based on individual number of lesions.

\section*{Results and Discussion}
Here we show our evaluation of the proposed BAGAU-Net on the 2017 MICCAI WMH segmentation challenge and the ABVIB datasets. A quantitative evaluation of the results is summarized in Table~\ref{wmh_result} and Table~\ref{abvib_result} respectively. Results showed that incorporate the knowledge of atlas can improve segmentation performance, and our proposed BAGAU-Net yields competitive results by achieving the best performance across all metric comparisons. More importantly, our proposed method has achieved better DSC and AVD scores compared to segmentation performance using standard FLAIR and T1-weighted images, which are important metrics for evaluating WMH segmentation in clinical images. The recall and F1 statistics were lower in the ABVIB data set than in the WMH Challenge data set. This was primarily due to the presence of a number of small WMH lesions in the ABVIB dataset that were not as prevalent in the WMH Challenge dataset. However the DSC and AVD metrics were similar for both data sets, suggesting that the same WMH lesion load for the whole brain was segmented correctly using the BAGAU-Net model. This was observed to be true in the visual review.

\begin{table}[!htbp]
\centering
\caption{Results of BAGAU-Net and other methods on WMH challenge datasets. $\downarrow$ indicates that the smaller the better (0=best, and 100=worst). $\uparrow$ indicates that the greater the better (0=worst, and 100=best).}
\label{wmh_result}

\begin{tabular}{lcccc}
\hline
Model & DSC (\%) $\uparrow$ & AVD (\%) $\downarrow$ & Recall (\%) $\uparrow$ & F1 (\%) $\uparrow$\\
\hline
U-Net \cite{sysu_media} (FLAIR + T1)& 81.39 & 16.63 & 81.41 & 78.12\\
U-Net \cite{sysu_media} (FLAIR) & 80.58 & 17.13 & 81.72 & 76.75\\
U-Net \cite{sysu_media} (FLAIR + Atlas) & 80.61 & 17.99 & 80.27 & 76.07\\
\hline
BAGAU-Net & \bfseries{82.02} & \bfseries{14.19} & \bfseries{82.58} & \bfseries{78.42}\\

\hline
\end{tabular}
\end{table}

\begin{table}[!htbp]
\centering
\caption{Results of BAGAU-Net and other methods on ABVIB datasets.}
\label{abvib_result}

\begin{tabular}{lcccc}
\hline
Model & DSC (\%) $\uparrow$ & AVD (\%) $\downarrow$ & Recall (\%) $\uparrow$ & F1 (\%) $\uparrow$\\

\hline
U-Net \cite{sysu_media} (FLAIR) & 77.48 & 21.18 & 51.50 & 56.13\\
U-Net  \cite{sysu_media} (FLAIR + Atlas) & 76.56 & 23.87 & 52.12 & 56.14\\

\hline
BAGAU-Net & \bfseries{80.27} & \bfseries{16.17} & \bfseries{58.51} & \bfseries{60.48}\\
\hline
\end{tabular}
\end{table}

\begin{figure*}[!htbp]
\centering
\subcaptionbox{Original}{\includegraphics[width=0.19\textwidth]{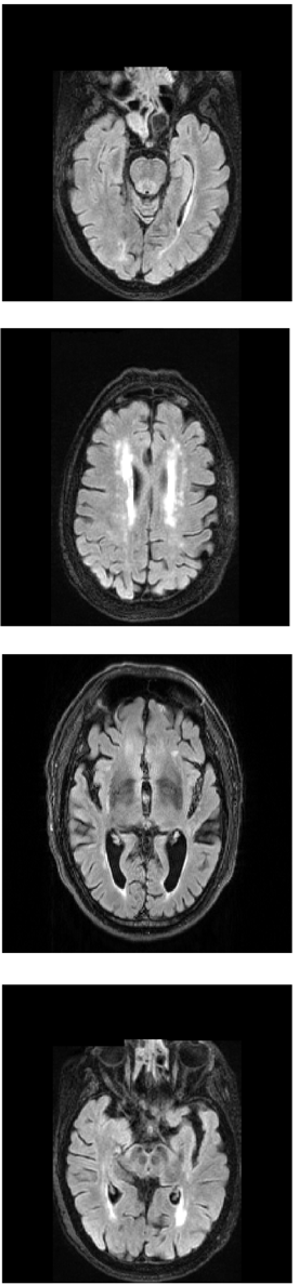}}
\hfill 
\subcaptionbox{Groud Truth}{\includegraphics[width=0.19\textwidth]{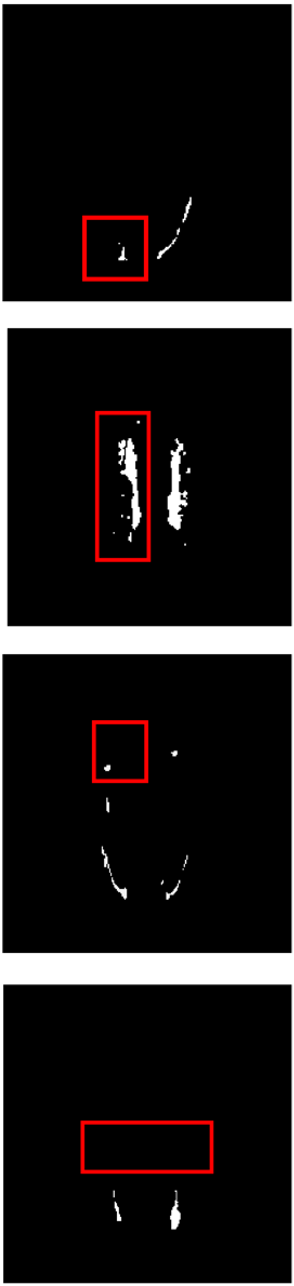}}%
\hfill 
\subcaptionbox{Flair + T1}{\includegraphics[width=0.19\textwidth]{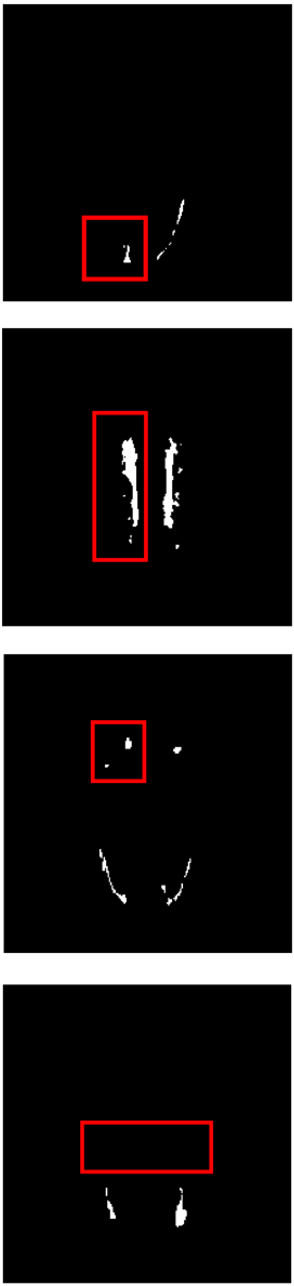}}%
\hfill
\subcaptionbox{Flair}{\includegraphics[width=0.19\textwidth]{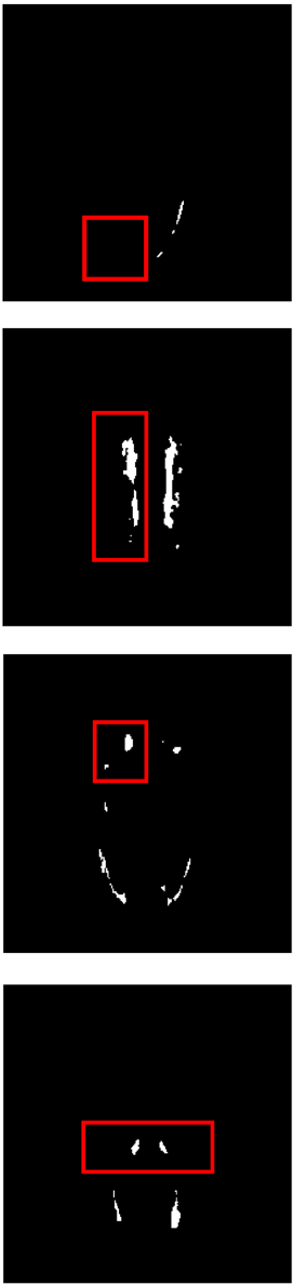}}%
\hfill 
\subcaptionbox{BAGAU-Net}{\includegraphics[width=0.19\textwidth, height=390pt]{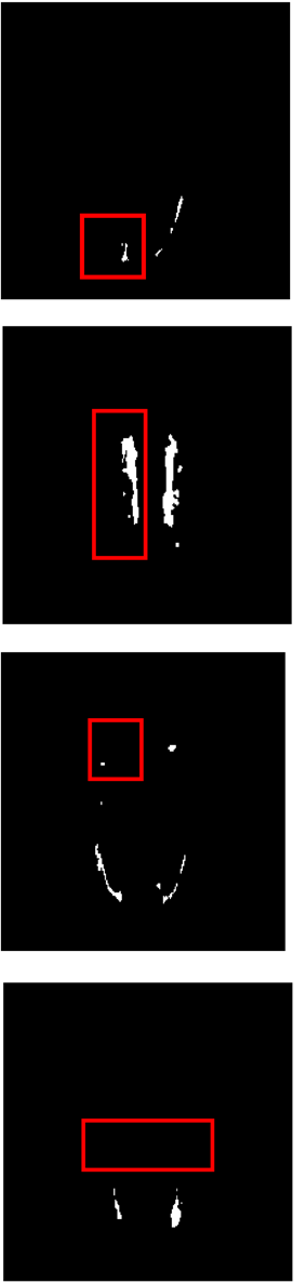}}%
\caption{Test images from the 2017 WMH challenge dataset (1st column), ground truth (2nd column), the segmentation results obtained by U-Net with FLAIR and T1 (3rd column), FLAIR only (4th) column), our proposed BAGAU-Net (5th column) with FLAIR and WM atlas.}
\label{result_compare}
\end{figure*}


Segmentation results are shown in Fig. \ref{result_compare}. It is clear that the absence of T1-weighted images can result in less accurate segmentation results as shown in column $4$ (FLAIR only) versus columns $3$ (FLAIR $+$ T1). It can be observed that BAGAU-Net, using only FLAIR and WM brain atlas, yield similar segmentation results compared to segmentation results using FLAIR and T1-weighted image as shown in Fig. \ref{result_compare} column $5$. Both experimental results and visual comparison showed that BAGAU-Net can effectively capture useful information from WM brain atlas and use it as prior guidance to yield more accurate segmentation results. Compared to the state-of-the-art method developed for WMH segmentation that requires the presence of T1-weighted image, our proposed method can provide even better results with only FLAIR and publicly available WM brain atlas. Furthermore, our model performs better compared to a recent review of the automated methods for WMH segmentation in which the calculated average DSC across studies was 0.73 \cite{t1}.

\begin{table}[!t]
\centering
\caption{Results of ablation study on MAM and AFM on WMH challenge dataset}
\label{ablation}

\begin{tabular}{lcccc}
\hline
Model & DSC (\%) $\uparrow$ & AVD (\%) $\downarrow$ & Recall (\%) $\uparrow$ & F1 (\%) $\uparrow$\\

\hline
BAGAU-Net (without MAM and AFM) & 80.52 & 17.33 & 81.64 & 74.19 \\
BAGAU-Net (without MAM)& 81.29 & 15.72 & 82.13 & 77.31 \\
BAGAU-Net (without AFM)& 80.93  & 16.19 & 81.87 & 76.63 \\

\hline
BAGAU-Net & \bfseries{82.02} & \bfseries{14.19} & \bfseries{82.58} & \bfseries{78.42}\\
\hline
\end{tabular}
\end{table}

To illustrate the effectiveness of different components in BAGAU-Net, we perform ablation studies on the WMH challenge datasets as shown in Table~\ref{ablation}. For our baseline method, we took out both MAM and AFM from the model and replace it with simple concatenation. Compared to our baseline, we observed that adding either attention modules improves segmentation performance across all metrics. Moreover, including both MAM and AFM further improves the results.

\subsection*{Model Limitation}

The main limitation of our model is its application in a relatively small sample size of patients with cSVD (total sample size is 90 subjects). However, more importantly, to address this limitation and increase the external validity of our model, our study subjects were derived from two different datasets (MICCAI and ABVIB). Furthermore, we have only utilized FLAIR sequences in this study to simulate real life scans in this study. The second limitation of our model is that we did not segment separately other lesion types of cSVD such as lacunes since lacunes may have different clinical and prognostic significance than WMH. Our aim in this study, however, was to first develop a successful paradigm for WMH since it is the most prevalent type of cSVD. Current paradigms are under development to subsequently segment lacunes within WMH regions. Finally, our current model was developed in cohorts of patients without other acute or chronic brain etiologies such as acute stroke or brain tumors which can introduce further heterogeneities in brain MRI scans in different clinical settings. The aim of this study, however, was to focus on the method development for combining brain atlas knowledge with deep learning for WMH alone. We will subsequently validate this approach in patients with other brain pathologies.  
\section*{Conclusion and Future Work}
In this paper, we proposed brain atlas guided attention U-Net to improve performance on WMH segmentation. Our model combined domain knowledge of WM brain atlas with deep convolution neural networks by introducing the segmentation path and the atlas encoding path. The proposed MAM and AFM mechanisms were applied at each decoding steps and final prediction step accordingly to help to capture more comprehensive features as well as increase model interpretability by incorporating prior knowledge. We showed that WMH segmentation with T1-weighted image could be replaced and even improved by using publicly available WM brain atlas with our proposed BAGAU-Net. Results showed that our proposed model has out-performed state-out-of-the-art for WMH segmentation on both datasets.
Datasets used in this study were collected from elderly group ($>$65), but the brain atlas we used were generated from a younger group (44$\pm$7) (currently, there is no public available brain atlas that focus on aging groups). Ventricles and sulci spaces in aging brains are generally larger than that of young brains \cite{atlas_survey}. Hence, our model can be further improved by the implementation of an aging white matter brain atlas.

\section*{Acknowledgements}
This work was funded in part by the National Science Foundation (grant number CBET-2037398) and the National Center for Advancing Translational Research (grant number UL1TR002733). The content is solely the responsibility of the authors and does not necessarily represent the official views of the funding agencies.

\makeatletter
\renewcommand{\@biblabel}[1]{\hfill #1.}
\makeatother

\bibliographystyle{unsrt}

\end{document}